\documentclass[10pt,letterpaper]{article}
\usepackage{opex3}

\usepackage{cite}
\usepackage{amsmath,amssymb,graphicx, color}

\DeclareGraphicsExtensions{.pdf, .png, .jpg}

\begin{document}

\title{Differential high-resolution stimulated CW Raman spectroscopy of hydrogen in a hollow-core fiber}

\author{Philip G. Westergaard,$^*$ Mikael Lassen, and Jan C. Petersen}

\address{Danish Fundamental Metrology, Matematiktorvet 307, DK-2800 Kgs. Lyngby, Denmark }

\email{$^*$Corresponding author: pgw@dfm.dk}

\begin{abstract}
We demonstrate sensitive high-resolution stimulated Raman measurements of hydrogen using a hollow-core photonic crystal fiber (HC-PCF). The Raman transition is pumped by a narrow linewidth ($<50\;$kHz) 1064 nm continuous-wave (CW) fiber laser. The probe light is produced by a homebuilt CW optical parametric oscillator (OPO), tunable from around 800 nm to 1300 nm (linewidth $\sim 5\;$MHz). These narrow linewidth lasers allow for an excellent spectral resolution of approximately $10^{-4}\;$cm$^{-1}$. The setup employs a differential measurement technique for noise rejection in the probe beam, which also eliminates background signals from the fiber. With the high sensitivity obtained, Raman signals were observed with only a few mW of optical power in both the pump and probe beams. This demonstration allows for high resolution Raman identification of molecules and quantification of Raman signal strengths.
\end{abstract}


\section{Introduction}
With applications in physical, chemical, biological sciences and in a number of different industrial areas \cite{Eckbrethbook1996, Sigrist2008, Hanf2014} there is growing demand for versatile and highly sensitive optical spectroscopy and microscopy tools. For this, Raman spectroscopy is a popular solution. Raman spectroscopy is based on molecular transitions between motional states and allows identification of molecules and materials \cite{Colthupbook1975}. Molecular vibrations reflect the molecular structure and are therefore used as a spectroscopic fingerprint for detection, identification, and imaging of a multitude of different molecules \cite{Freudiger2008, Das2011}. Traditionally, the Raman scattered (Stokes) signal has been obtained via a spontaneous process requiring high optical intensity in the pump laser beam, based on possibly expensive pulsed laser sources which require elaborate techniques for hyperspectral (i.e. multiple molecules) detections. Recently, more focus has been given to a different approach were the Raman transition is stimulated coherently using two lasers with different frequencies, with the frequency difference between the two lasers corresponding to the energy difference in the Raman transition involved \cite{Colthupbook1975}. Stimulating the transition in this way instead of relying on a spontaneous process gives orders of magnitude larger signal \cite{BenabidScience2002} and makes it possible to observe the Raman signal with CW lasers with moderate optical power in the sub-watt range, which otherwise for spontaneous excitation would require Watt levels of CW power in the best case \cite{BenabidPRL2007, Couny2009}. In further pursuit of higher Raman signal-to-noise ratio, efforts have been made to contain molecules inside a hollow-core fiber \cite{BenabidNature2005, RussellNatPhot2014, Russell2003}. This ensures high intensity over a long interaction length between the light and the molecules. It has been shown that hollow-core photonic crystal fibers (HC-PCFs) are very efficient for enhancing the Raman signals for both spontaneous and stimulated Raman transitions \cite{Buric2008, Dom2013}. For stimulated Raman spectroscopy two lasers are used. When the frequency difference between the two lasers matches that of an allowed Raman transition, the pump laser (typically the high frequency laser) will be depleted, while the probe laser (typically the low frequency laser) will experience a gain in intensity. This can be seen from the following expression for the change induced in the optical probe power ($\delta P(\nu_{probe})$) \cite{Esherick1982, Dom2013},
\begin{equation}
\delta P(\nu_{probe})= \frac{N \pi}{h c^2 \nu_{probe}^{2}} \left( \frac{d^2\sigma}{d\Omega d\nu_{probe}} \right) P(\nu_{pump}) P(\nu_{probe}),
 \label{eq.gain}
\end{equation}
where $N$ is the population difference between upper and lower molecular states, $\left( \frac{d^2\sigma}{d\Omega d\nu_{probe}} \right)$ is the spectrally resolved differential Raman scattering cross section and $P(\nu_{pump}) \left( P(\nu_{probe}) \right)$ is the optical power of the pump (probe) beam. Equation (\ref{eq.gain}) shows that the change in the optical probe power is proportional to the product of the pump and probe powers $\left( P(\nu_{pump})P(\nu_{probe}) \right) $ and proportional to the number of molecules through the population difference $N$ under constant thermal conditions. The measurements presented here confirm this functional dependence experimentally.

In this work, measurements are performed on the S$_0$(1) transition of ortho-hydrogen at $588 \;$cm$^{-1}$, which is the most intense rotational line at room temperature for lower wavenumbers. We demonstrate a setup with stimulated Raman spectroscopy (SRS) inside a hydrogen filled hollow-core fiber using two narrow linewidth CW lasers; a Koheras BoostiK fiber laser at 1064 nm (linewidth $<50$ kHz) and a homebuilt OPO at 1135 nm (linewidth $\sim 5$ MHz). Using these two narrow linewidth lasers ensures that spectral features in principle can be measured with a very high instrumental resolution of around 5 MHz corresponding to $1.6\times 10^{-4}$ cm$^{-1}$. However, in our regime the actual width of the Raman transition will be limited by other effects such as Doppler broadening, wall collisions or transition‐time broadening, and pressure broadening. The signal-to-noise ratio (SNR) is enhanced by using modulation detection such as shown in \cite{Dom2013}. In this work, the 1064 nm pump laser is frequency modulated and this modulation is transferred to the probe laser via the Raman transition and demodulated with a lock-in amplifier giving the Raman signal. However, compared to \cite{Dom2013} we have further improved the setup by employing differential measurement detection of the probe laser, while suppressing the pump laser. With this technique any classical noise in the probe beam can be rejected \cite{Wang}, thus enhancing the SNR of the Raman signals. This setup ensures a high SNR and has allowed us to quantify the contributing parameters (optical power and pressure) with high precision.

\section{Experimental setup}
\begin{figure}[ht]
\centering\includegraphics[width=1.0\columnwidth]{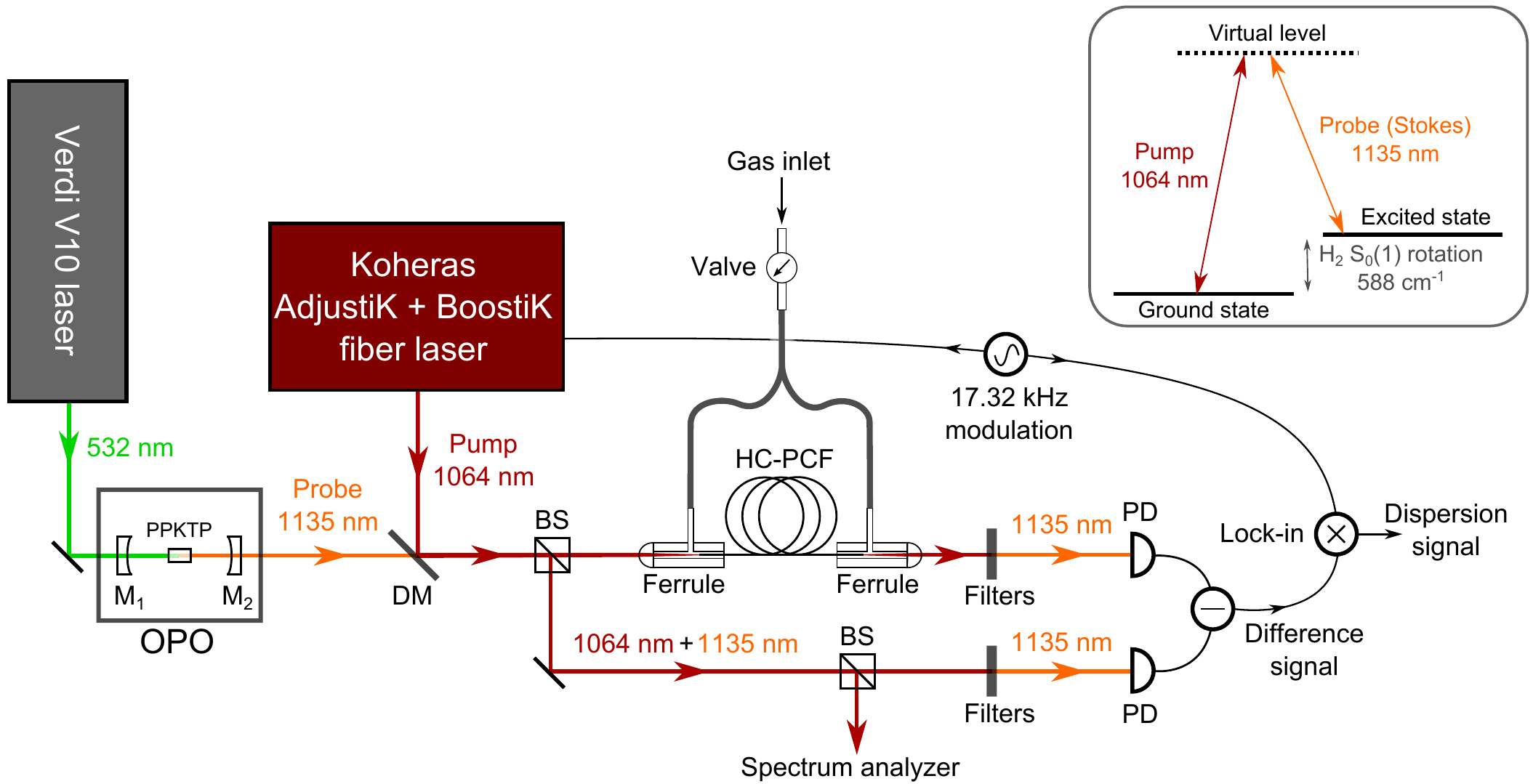}
\caption{An overview of the experimental setup. DM: Dichroic Mirror, BS: Beam Splitter, PD: Photo Diode. M$_{1,2}$: OPO mirrors. The inset shows a schematic of the Raman transition studied here.}
\label{Fig:Setup}
\end{figure}
An overview of the experimental setup is shown in Fig.~\ref{Fig:Setup}. The probe light at 1135 nm is generated using an OPO with a 20 mm long PPKTP crystal pumped at 532 nm by a Verdi V10 laser. The OPO cavity is 55 mm long, linear, and consists of two concave mirrors M$_{1,2}$ with ROC = 50 mm, where M$_1$ reflects 30$\%$ and 95$\%$ of the incident power for 532 nm and 1135 nm, respectively, while M$_2$ reflects 99.9$\%$ of the power for both 532 nm and 1135 nm. 
The OPO has a linewidth at 1135 nm of $\sim 5$ MHz and the threshold pump power was measured to be 25 mW. The OPO can be tuned in the wavelength range from approximately 800 nm to 1300 nm \cite{Wiese2000} by tuning the temperature of the PPKTP crystal and can provide 10-15 mW of optical power at 1135 nm. An optical spectrum analyzer is used for monitoring the wavelength of the OPO. This is especially important here since thermal heating of the crystal can change the wavelength. The OPO cavity is locked to the Verdi laser frequency using the tilt locking technique \cite{Shaddock1999} (not shown in Fig. ~\ref{Fig:Setup}). The probe light is combined on a dichroic mirror with the pump light at 1064 nm. The 1064 nm fiber pump laser has a linewidth of $< 50$ kHz and can be tuned approximately 0.9 nm by changing the temperature of the laser. Using these narrow linewidth lasers we have the capability of performing measurements with extremely high spectral resolution on the Raman transitions and in principle the setup is able to resolve features down to around 5 MHz (or $1.6 \times 10^{-4}$ cm$^{-1}$). After combining the probe and pump beams on the dichroic mirror, the light is split into two arms where one beam is coupled through a HC-PCF and the other beam is used as the reference for the differential measurement. The HC-PCF is a 4.5 m long piece of custom made fiber from NKT Photonics. The fiber shares most characteristics with the standard HC-1060-02 fiber \cite{NKT}, such as structure, hole diameter of 10 $\mu$m, polarization properties, expect that the transmission window for this fiber was extended to 940-1170 nm. The fiber is filled with hydrogen gas at variable pressures up to about 1 atm. We use special in-house developed ferrules that allow for a compact solution for both coupling of the light and gas filling into the fiber in the same device. The HC-PCF is inserted in the back of the ferrule and the gap is sealed with vacuum grease. The ferrules have a lens glued to the front face for optical coupling and a glass tube attached to the side for gas filling. The best coupling efficiency into the HC-PCF using these ferrules was approximately 40$\%$ for both probe and pump beams.

\section{Noise rejection and enhancement of SNR}
The pump laser (1064 nm) is frequency modulated at 17.32 kHz which is close to the highest possible frequency for this laser. This modulation is imprinted onto the wavelength of the probe laser (1135 nm) inside the HC-PCF only if the Raman transition takes place. The 1064 nm light is blocked by a factor of more than $10^4$ by filters placed in front the differential detector, and only the probe light (1135 nm) reaches the two photo diodes of the differential detector.
The signals from the photo diodes are electronically subtracted before being de-modulated at 17.32 kHz using a lock-in amplifier and sampled using a 12 bit oscilloscope. The differential detection ensures that common mode noise in the probe is rejected by subtracting the signal from the beam passing through the HC-PCF and the signal from the reference beam in an interferometer-like configuration, followed by demodulation of the signal at the pump modulation frequency. This results in a low-noise signal that will be detected only when the Raman transition occurs.

\begin{figure}[h]
\centering\includegraphics[width=10cm]{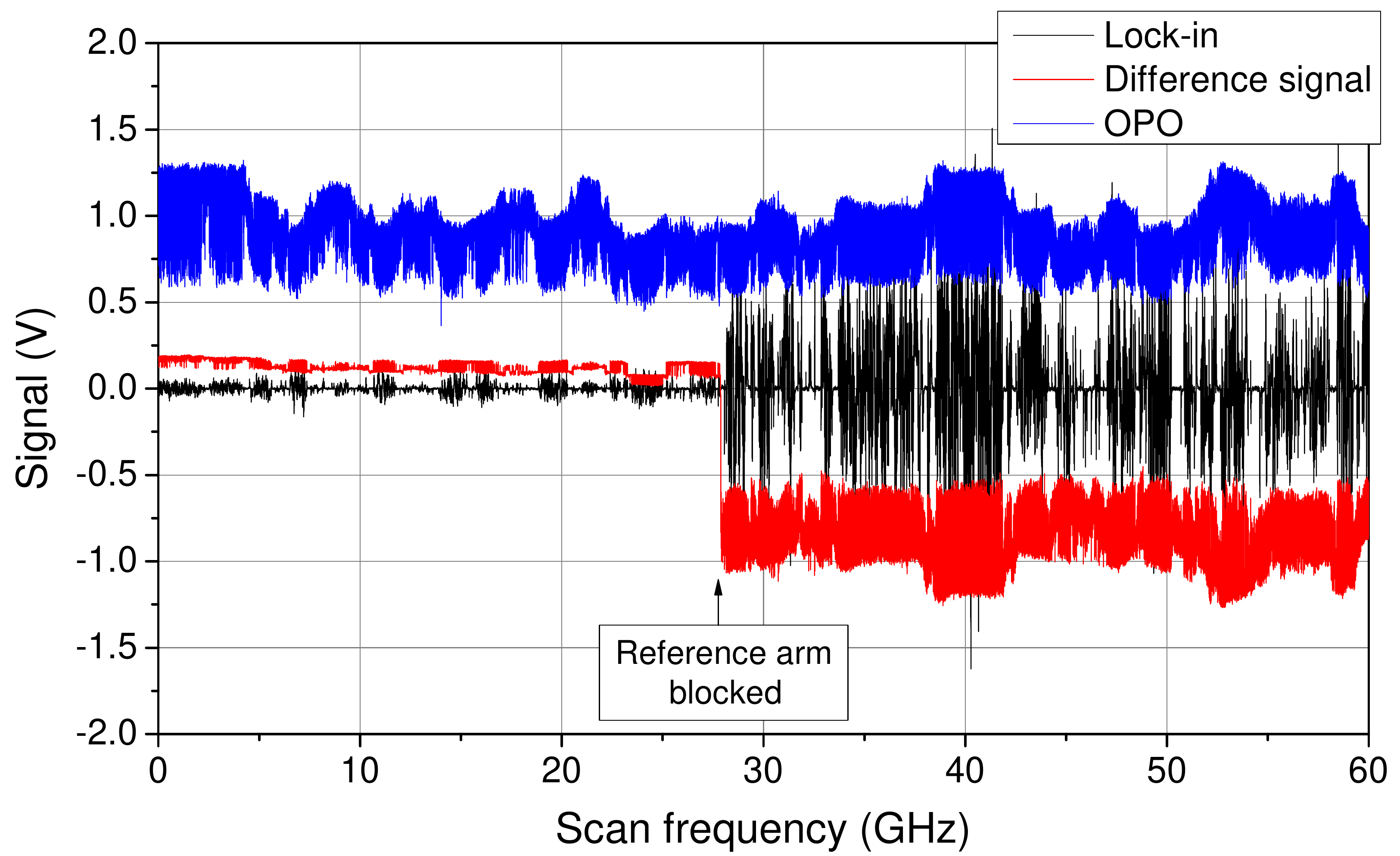}
\caption{Effect of the differential detection scheme. The blue curve shows the OPO output power measured with a DC photo detector. The red curve shows the difference signal and the black curve is the lock-in signal.  The reference arm is blocked during a scan resulting in increased noise in both the detector and lock-in signals. All measurements are made with an integration time of 30 ms for the lock-in amplifier. Each trace contains 100000 data points.}
\label{Fig:NoiseRejection}
\end{figure}

We use a very short integration time of 30 ms for the lock-in amplifier in order to have the SNR of the Raman signal without too much averaging. The differential detector subtracts the DC power of the probe to avoid saturation of the lock-in amplifier, and any background signal from spontaneous Raman scattering of the pump light from silica in the fibre, which could otherwise be problematic \cite{Hanf2014}, only leaves a small constant offset in the lock-in signal which can be subtracted in post-processing.

\begin{figure}[h]
\centering\includegraphics[width=13cm]{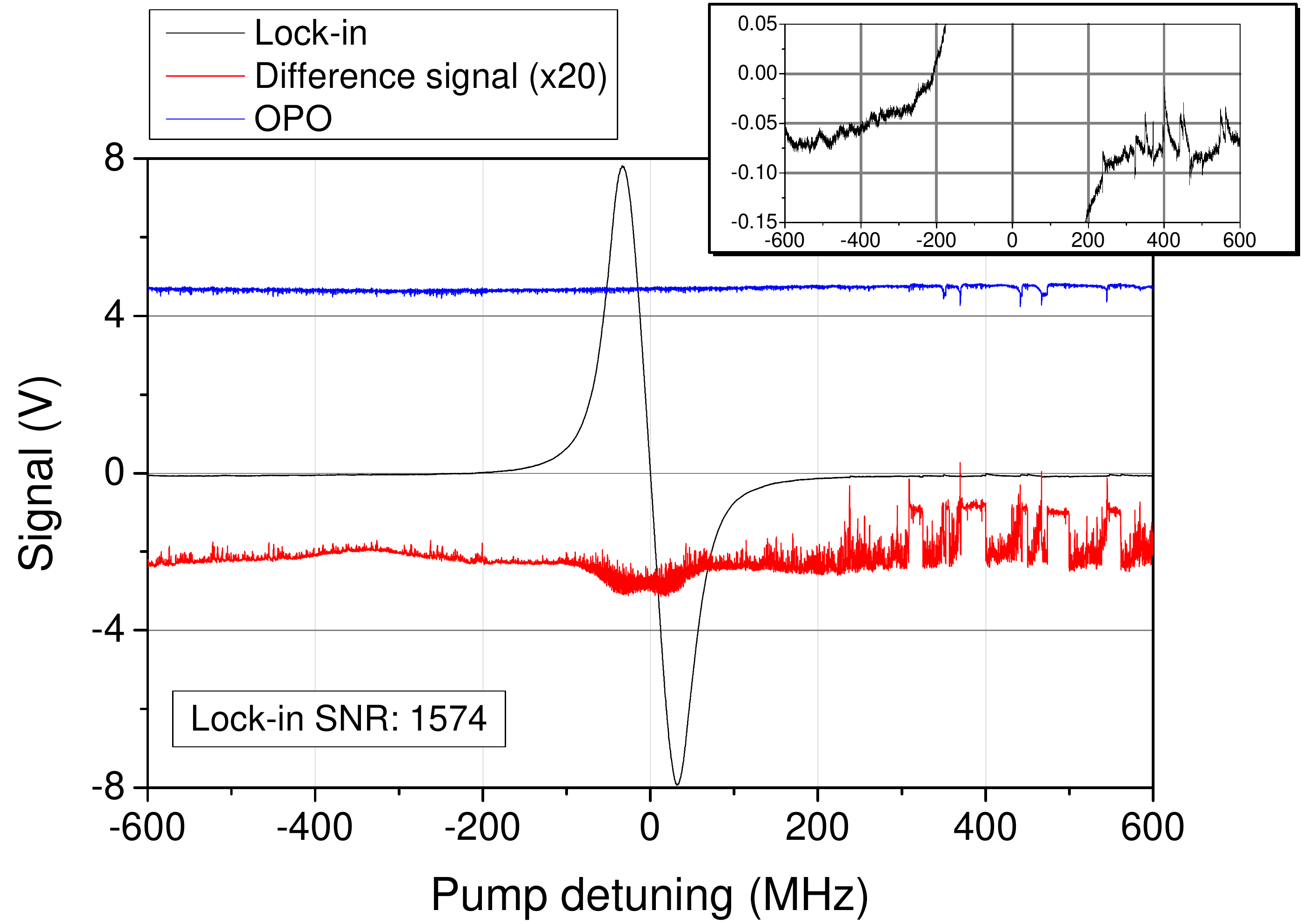}
\caption{Stimulated Raman signal from H$_2$ as a function of the pump laser frequency relative to the Raman transition displaying the signals involved and the noise reduction from both the electronic subtraction and lock-in detection. The blue curve shows the OPO output power, the red curve shows the difference signal, and the black curve is the lock-in signal. The H$_2$ pressure in the fiber was 867 hPa, the transmitted power of the probe (OPO) light was 1.85 mW and the transmitted power of the pump light was 230 mW. The SNR is approximately 1600. The inset shows a blow-up of the vertical scale for the lock-in signal with the same units on both axes. The noise suppression from the differential detection remains around 20 dB here. The frequency axis was calibrated with a Fabry-P\'{e}rot cavity to an accuracy of 1 MHz.}
\label{Fig:SNR}
\end{figure}

Figure~\ref{Fig:NoiseRejection} shows the effect of the differential detection scheme. The pump laser frequency was scanned off-resonance and at one point the reference arm was blocked to demonstrate the increase in noise. This would be the normal level of noise without the differential detection scheme. We find that the standard deviation of the lock-in signal is 0.025 V and 0.28 V in the case of differential detection and blocked reference arm (without differential detection), respectively. The differential scheme thus reduces the noise of the detected signal by a factor of 10 ($\sim 20$ dB). Note that in order to make these measurements the OPO was allowed to run freely (i.e., the cavity was not locked to the 532 nm laser), only with thermal self-locking and therefore the intensity noise is relatively high. This can clearly be seen by comparing the blue curve in Fig.~\ref{Fig:NoiseRejection} and the blue curve in Fig.~\ref{Fig:SNR}, where the OPO cavity was more tightly locked.

Figure~\ref{Fig:SNR} shows a scan over the Raman transition that also demonstrates the reduction in noise from the different stages in the detection system, as detailed in the following. The OPO is locked to the 532 nm light and the output signal (blue curve) shows a photo diode signal of the optical power of the probe laser. The power transmitted through the HC-PCF was about $1.8$ mW. The small dips in the signal (at detunings of around $350-550$ MHz) are due to mode jumps of the OPO frequency. These occur with varying intervals due to thermal effects in the OPO cavity, and can be reduced or eliminated with a different choice of cavity mirrors, which on the other hand would increase the lasing threshold of the OPO and also increase the linewidth. The effect of these mode jumps are still visible in the difference signal (red curve, amplified by a factor 20), but less dominant compared to the noise level of the signal. Finally, the de-modulated lock-in signal (black curve) produces the dispersion quadrature of the Raman transition. The off-resonant noise is reduced by a large factor, resulting in a SNR of 1574. The mode-jumps are still the dominant source of noise, but they do not disturb the resonance profile significantly. Note that the SNR at this point is limited by the electrical noise of the optical detector, and the achievable SNR can be much higher. Nonetheless, the SNR indicates that Raman signals can be detected with this setup using only a few (mW)$^2$ in the power product between the pump and probe beams at the pressure indicated. On the other hand, keeping the power product constant the SNR also indicates that we can measure Raman signals at very low pressures, down to 1-2 hPa. This scaling of the Raman signal with the power product (mW)$^2$ and pressure will be addressed in the next section.

\section{Quantitative results}
With the sensitive detection system at hand, we have made a quantitative analysis of the system's parameters and tested the linearity of our system. The stimulated Raman signal scales with the product of the pump and probe power (see Eq. \eqref{eq.gain}), and thus scales linearly with pump power if the probe power is held constant.
\begin{figure}[!h]
\centering\includegraphics[width=10cm]{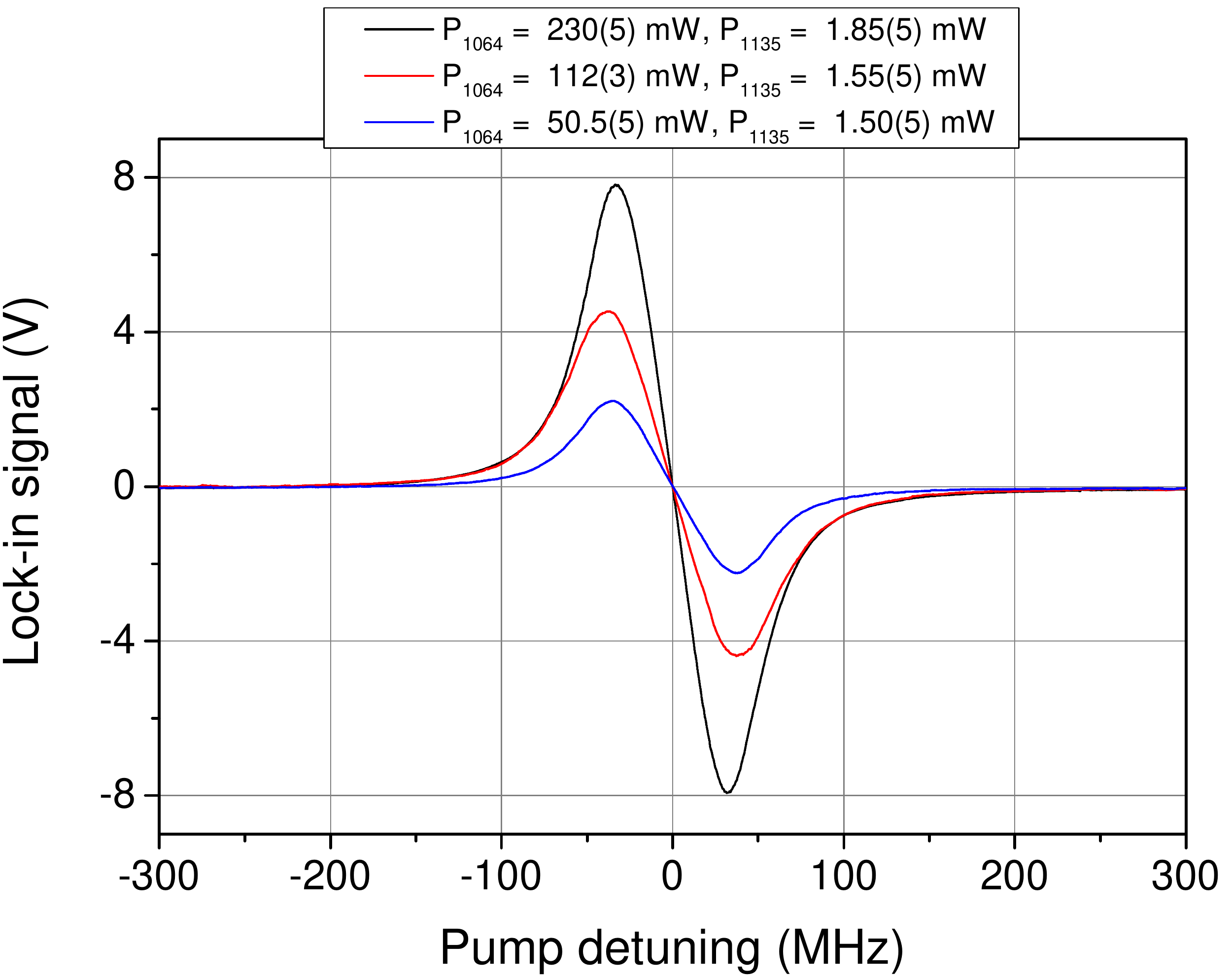}
\caption{A scan over the Raman transition for different pump powers with the probe power roughly constant (at the level 1.5-1.8 mW). The signal amplitude scales linearly with pump power. The H$_2$ pressure in the fiber was 867 hPa. Numbers in parenthesis denote uncertainty of the final digit corresponding to one standard deviation.}
\label{Fig:power}
\end{figure}
In Fig.~\ref{Fig:power} we show scans over the Raman transition with varying optical power of the pump laser. As expected, the signal amplitude scales linearly with pump power and we are able to see clear spectral features down to a few (mW)$^2$ of power in the product between the pump and probe beams, specifically we have observed the Raman signal with 1.1 mW probe power and 2.1 mW pump power. Since the Raman signals scale directly with the product of the pump and probe optical powers (see Eq.(\ref{eq.gain})) we can increase the SNR simply by increasing the optical power in either the pump or probe or both. However, one has to consider the damage threshold of the fiber and nonlinear effects that might occur at very high optical powers.

\begin{figure}[!h]
\centering\includegraphics[width=10cm]{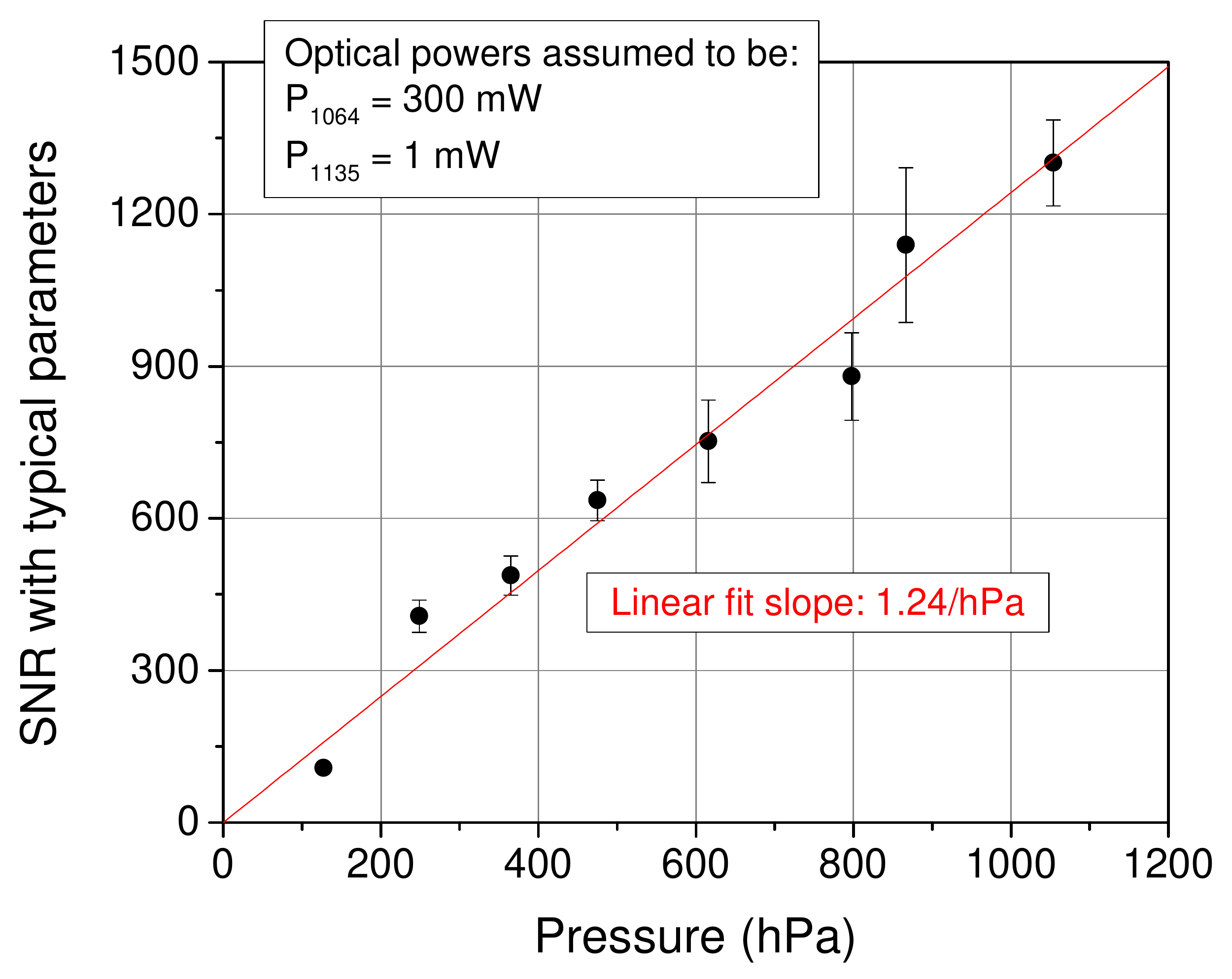}
\caption{The obtainable signal-to-noise ratio as a function of H$_2$ pressure inside the HC-PCF with typical parameters for the pump power (P$_{1064} = 300\;$mW) and the probe power (P$_{1135} = 1$ mW). The red line is a linear fit to data with the intercept set to zero.}
\label{Fig:pressure}
\end{figure}

Figure~\ref{Fig:pressure} shows the scaling of the SNR with pressure inside the HC-PCF. We have estimated the filling time of the fiber from empty to ambient pressure to around a few hours using the Knudsen´s equation \cite{Henningsen2008}. This takes into account the increase in filling speed by filling the fiber from both ends. After changing the pressure we wait for at least 4 hours before conducting new measurements. The data presented in Fig.~\ref{Fig:pressure} have been normalized to a given optical power in the pump and probe beams for each measurement and all the data are scaled to the same typical powers for comparison. A linear fit to the data gives a slope in the SNR of $(1.24\pm 0.04)$/hPa, and it would thus be possible to detect the H$_2$ Raman transition at pressures down to a few hPa with the optical power levels in the current system.

Additionally, from measurements at six different pressures ranging from 128 hPa to 1054 hPa we find the FWHM linewidth of the Raman transition to be $(81.2\pm 1.9)\;$MHz, where the value is the average of the individual measurements weighted with their inverse uncertainty and the error bar is the standard deviation of this average. The linewidth does not show any pressure dependence over the whole pressure range within the error bars of the individual measurements. This could be attributed to different effects such as collisional narrowing, pressure and wall broadening \cite{Murray1972, Owyoung1978}. Compared to previous results \cite{Flohic} however, the lack of pressure dependence is quite surprising and is something we are currently investigating. 

\newpage

\section{Conclusion}
We have demonstrated the use of differential detection resulting in superior sensitivity of CW stimulated Raman scattering with high spectral resolution from hydrogen gas inside a hollow-core fiber. It was demonstrated that for a few hundred mW of pump power the SNR was approximately 1600, which currently is limited by the electronic noise of the detection system. We found that the differential scheme reduces the noise of the detected signal by a factor of 10 $( 20 \textrm{ dB})$. We have characterized our differential detector and found that the electronic subtraction delivers at least 25 dBs of classical noise suppression. When the optical power is increased to more than about 12 mW in each detector, the differential noise suppression starts to saturate and any classical noise present in the light will start to contribute to the overall noise level. Up to a probe power of 12 mW the detector thus removes classical noise from the probe, and we would therefore gain at least a factor of 6 in the SNR, if the probe power was increased from the current $\sim 1.8$ mW to 12 mW. Furthermore, an additional factor of 10 in SNR can be gained by increasing the pump power to 2-3 W without any damage to the fiber \cite{BenabidPRL2007}.

The system presented here is in principle able to measure Raman transitions with only a few (mW)$^2$ of power product between the pump and probe beams at ambient pressure. For a power product of $300$ (mW)$^2$ Raman signals can be measured at pressures down to a few hPa. Furthermore, by using these very narrow lasers we can in principle resolve features of Raman transitions down to around 5 MHz or $1.6\times 10^{-4}$ cm$^{-1}$. This high resolution allowed an ultra-precise measurement of the hydrogen $S_0(1)$ Raman transition linewidth of $(81.2\pm 1.9)\;$MHz. Future work will involve an investigation of the dependence of the Raman transition linewidth on pressure down to a few hPa. 

The presented setup is not limited to hydrogen or the specific energy range examined here since the OPO is tunable from around 800 nm to 1300 nm. This makes the system very useful for the measurements of multiple Raman transitions since the high resolution allows to easily distinguish between different Raman peaks. We believe that the SNR can be further enhanced by improving the OPO power stability and increasing the optical power, and by using an optical detector with lower electrical noise. 

\section*{Acknowledgments}
We thank NKT Photonics for supplying the HC-PCF, Jan W. Thomsen for loan of the Verdi laser, and Marco Triches for producing the ferrules.


\begin{thebibliography}{99}

\bibitem{Eckbrethbook1996} A. C. Eckbreth, \emph{Laser Diagnostics for Combustion Temperature and Species} (Gordon and Breach Publishers, 1996).

\bibitem{Sigrist2008} M. W. Sigrist, R. Bartlome, D. Marinov, J. M. Rey, D. E. Vogler, and H. W\"{a}chter, "Trace gas monitoring with infrared laser-based detection schemes," Appl. Phys. B \textbf{90}, 289--300 (2008).

\bibitem{Hanf2014} S. Hanf, T. B\"{o}g\"{o}zi, R. Keiner, Torsten Frosch, and J\"{u}rgen Popp, "Fast and highly sensitive fiber-enhanced Raman spectroscopic monitoring of molecular H2 and CH4 for point-of-care diagnosis of malabsorption disorders in exhaled human breath," Anal. Chem. \textbf{87}, 982--988 (2015).

\bibitem{Colthupbook1975}  N. B. Colthup, L. H. Daly, and S. E. Wiberley, \emph{Introduction to Infrared and Raman Spectroscopy} (Academic, 1975).

\bibitem{Freudiger2008} C. W. Freudiger, W. Min, B. G. Saar, S. Lu, G. R. Holtom, C. He, J. C. Tsai, J. X. Kang, and X. S. Xie, "Label-free biomedical imaging with high sensitivity by stimulated Raman scattering microscopy," Science \textbf{322}, 1857--1861 (2008).

\bibitem{Das2011} R. S. Das and Y. K. Agrawal, "Raman spectroscopy: recent advancements, techniques and applications," Vib. Spectrosc. \textbf{57}, 163--176 (2011).

\bibitem{BenabidScience2002} F. Benabid, J. C. Knight, G. Antonopoulos, and P. St. J. Russell, "Stimulated Raman scattering in hydrogen-filled follow-core photonic crystal fiber," Science \textbf{298}, 399--402 (2002).

\bibitem{BenabidPRL2007} F. Couny, F. Benabid, and P. S. Light, "Subwatt threshold cw Raman fiber-gas laser based on H$_2$-filled hollow-core photonic crystal fiber," Phys. Rev. Lett. \textbf{99}, 143903--143907 (2007).

\bibitem{Couny2009} F. Couny, O. Carraz, and F. Benabid, "Control of transient regime of stimulated Raman scattering using hollow-core PCF," J. Opt. Soc. Am. B \textbf{26}, 1209--1215 (2009).

\bibitem{BenabidNature2005} F. Benabid, F. Couny, J. C. Knight, T. A. Birks, and P. St J. Russell, "Compact, stable and efficient all-fibre gas cells using hollow-core photonic crystal fibres," Nature \textbf{434}, 488--491 (2005).

\bibitem{RussellNatPhot2014} P. St. J. Russell, P. H\"{o}lzer, W. Chang, A. Abdolvand, and J. C. Travers, "Hollow-core photonic crystal fibres for gas-based nonlinear optics," Nat. Photonics \textbf{8}, 278--286 (2014).

\bibitem{Russell2003} P. St. J. Russell, "Photonic crystal fibers," Science \textbf{299}, 358--362 (2003).

\bibitem{Buric2008} M. P. Buric, K. P. Chen, J. Falk, and S. D. Woodruff, "Enhanced spontaneous Raman scattering and gas composition analysis using a photonic crystal fiber," Appl. Optics \textbf{47}, 4255--4261 (2008).

\bibitem{Dom2013} J. L. Dom\'{e}nech and M. Cueto, "Sensitivity enhancement in high resolution stimulated Raman spectroscopy of gases with hollow-core photonic crystal fibers", Opt. Lett. \textbf{38}, 4074--4077 (2013).

\bibitem{Esherick1982} P. Esherick and A. Owyoung in \emph{Advances in Infrared and Raman Spectroscopy}, R. J. H. Clark and R. E. Hester, eds. (Heyden, 1982) pp. 130–-187.

\bibitem{Wang}	Q. Wang, J. Chang, C. Zhu, Y. Liu, G. Lv, F. Wang, X. Liu, and Z. Wang, "High-sensitive measurement of water vapor: shot-noise level performance via a noise canceller," Appl. Optics \textbf{52}, 1094--1099 (2013).

\bibitem{Wiese2000} D. R. Wiese, U. Stroessner, A. Peters, J. Mlynek, S. Schiller, A. Arie, A. Skliar, G. Rosenman, "Continuous-wave 532-nm-pumped singly resonant optical parametric oscillator with periodically poled KTiOPO$_4$," Opt. Comm. \textbf{184}, 329--333 (2000).

\bibitem{Shaddock1999} D. A. Shaddock, M. B. Gray, and D. E. McClelland, "Frequency locking a laser to an optical cavity by use of spatial mode interference," Opt. Lett. \textbf{24}, 1499--1501 (1999).

\bibitem{NKT} NKT Photonics A/S, "HC-1060-02," www.nktphotonics.com/files/files/HC-1060.pdf 

\bibitem{Henningsen2008} J. Henningsen and J. Hald, "Dynamics of gas flow in hollow core photonic bandgap fibers," Appl. Optics \textbf{47}, 2790--2797 (2008).

\bibitem{Murray1972} J. Murray, and A. Javan, "Effects of collisions on Raman line profiles of hydrogen and deuterium gas," J. Mol. Spectrosc. \textbf{42}, 1--26 (1972).

\bibitem{Owyoung1978} A. Owyoung, "High-resolution cw stimulated Raman spectroscopy in molecular hydrogen," Opt. Lett. \textbf{2}, 91-–93 (1978).

\bibitem{Flohic} M. P. Le Flohic, P. Duggan, P. M. Sinclair, J. R. Drummond, and A. D. May, "Collisional broadening and shifting of the pure rotational Raman lines S$_0$(J = 0–4) of H$_2$ at room temperature," Can. J. Phys. \textbf{72}, 186--192 (1994).

\end{thebibliography}
\end{document}